\documentclass[apj]{emulateapj}
\pdfoutput=1
\newcommand{\myion}[2]{\ensuremath{\mathrm{#1}^{#2}}}
\newcommand{\um}{\,\micron}
\newcommand{\arII}{[Ar\,\textsc{ii}]}
\newcommand{\arIII}{[Ar\,\textsc{iii}]}
\newcommand{\neII}{[Ne\,\textsc{ii}]}
\newcommand{\neIII}{[Ne\,\textsc{iii}]}
\newcommand{\neV}{[Ne\,\textsc{v}]}
\newcommand{\sIII}{[S\,\textsc{iii}]}
\newcommand{\sIV}{[S\,\textsc{iv}]}
\newcommand{\siII}{[Si\,\textsc{ii}]}
\newcommand{\siXIII}{Si\,\textsc{xiii}}
\newcommand{\siXIV}{Si\,\textsc{xiv}}

\newcommand{\oIV}{[O\,\textsc{iv}]}
\newcommand{\feII}{[Fe\,\textsc{ii}]}
\newcommand{\niII}{[Ni\,\textsc{ii}]}
\newcommand{\feIII}{[Fe\,\textsc{iii}]}

\shortauthors{Smith \textit{et al.}}  
\shorttitle{Cassiopeia A}

\begin{document}

\title{Spitzer Spectral Mapping of Supernova Remnant Cassiopeia A}

\author{%
J.D.T. Smith\altaffilmark{1,2}, 
Lawrence Rudnick\altaffilmark{3},
Tracey Delaney\altaffilmark{4},
Jeonghee Rho\altaffilmark{5},
Haley Gomez\altaffilmark{6},
Takashi Kozasa\altaffilmark{7},
William Reach\altaffilmark{5},
and Karl Isensee\altaffilmark{3}
}

\email{jd.smith@utoledo.edu}

\altaffiltext{1}{Ritter Astrophysical Observatory, University of Toledo,
  Toledo, OH 43606}

\altaffiltext{2}{Steward Observatory, University of Arizona, Tucson, AZ
 85721} 

\altaffiltext{3}{Astronomy Department, University of Minnesota,
 Minneapolis, MN 55455}

\altaffiltext{4}{Massachusetts Institute of Technology, Kavli Institute
 for Astrophysics and Space Research, Cambridge, MA 02139}

\altaffiltext{5}{Spitzer Science Center, Caltech, Pasadena, CA 91125}

\altaffiltext{6}{School of Physics and Astronomy, Cardiff University,
  Queens Buildings, The Parade, UK, CF24 3AA}

\altaffiltext{7}{Department of Cosmosciences, Graduate School of
  Science, Hokkaido University, Sapporo 060-0810, Japan}

\begin{abstract}
  We present the global distribution of fine structure infrared line
  emission in the Cassiopeia A supernova remnant using data from the
  Spitzer Space Telescope's Infrared Spectrograph.  We identify
  emission from ejecta materials in the interior, prior to their
  encounter with the reverse shock, as well as from the post-shock
  bright ring.  The global electron density increases by $\gtrsim$\,100
  at the shock to $\sim\!10^4$\,cm$^{-3}$, providing evidence for strong
  radiative cooling. There is also a dramatic change in ionization state
  at the shock, with the fading of emission from low ionization interior
  species like \siII\, giving way to \sIV\ and, at even further
  distances, high-energy X-rays from hydrogenic silicon.  Two compact,
  crescent-shaped clumps with highly enhanced neon abundance are
  arranged symmetrically around the central neutron star.  These neon
  crescents are very closely aligned with the ``kick'' direction of the
  compact object from the remnant's expansion center, tracing a new axis
  of explosion asymmetry.  They indicate that much of the apparent
  macroscopic elemental mixing may arise from different compositional
  layers of ejecta now passing through the reverse shock along different
  directions.
\end{abstract}


\section{Introduction}

Supernova remnant Cassiopeia A is both nearby (3.4kpc) and very young
\citep[$\sim$330yr,][]{Fesen2006}, giving it a bright, richly detailed
ejecta structure which has led to intensive study at many wavelengths,
from gamma rays to radio \citep{Rudnick2002,Albert2007}.  Spectroscopy
of distant light echoes from the original blast indicates that Cas A was
the type IIb core-collapse of a $\sim$15\,M$_\sun$ main sequence star
\citep{Krause2008}, which models indicate are highly stratified, having
lost a significant portion of the original outer layers of hydrogen and
helium prior to collapse in stellar winds \citep{Woosley1987}.  An
initial outgoing ``forward'' shock wave driven by the blast is seen as a
thin X-ray edge \citep{Gotthelf2001} expanding at $\sim$5000\,km/s, with
a reverse shock being driven back into the outgoing ejecta as it
interacts with circumstellar material and the interstellar medium,
expanding at roughly half the rate of the forward shock
\citep{DeLaney2003}.

The high resolution available (1\arcsec=0.016pc) makes Cas A ideal for
tracking the physical conditions in the layered ejecta material,
following nucleosynthesis both before and during the supernova
explosion, the subsequent mixing and inhomogenous transport of the
ejected materials, and the formation of dust \emph{in situ}.  The
mid-infrared (MIR) is an ideal wavelength for studying these processes,
as it is minimally affected by the significant line of sight dust
extinction \citep[$A_V=$ 4.6 -- 6.2,][]{Hurford1996}, and probes a rich
suite of abundant ionized species including Si, S, Ne, and Ar which
range in ionization energies from 8 -- 100eV.  Previous MIR studies of
the remnant using Infrared Space Observatory (ISO) mapping placed
constraints on the dust formation and the appearance of macroscopic
mixing of materials \citep{Lagage1996,Douvion1999}, and explored the
velocity shifted line and dust emission in various regions using
high-resolution MIR spectroscopy\citep{Arendt1999}.

We report on the first large, sensitive infrared spectral maps of the
remnant, obtained by the Spitzer Space Telescope \citep{Werner2004}. The
distribution of interstellar material, ejecta, and shocked circumstellar
material is shown in the 24\um\ maps of \citet{Hines2004}.  A model of
the composition and mass of emitting dust associated with the remnant
using this MIR data set was undertaken by \citet{2008ApJ...673..271R}.
\citet{2006ApJ...652..376E} highlighted the associated deep 3.6--8\um\
Spitzer IRAC imaging, exploring the survival of the various nuclear
burning layers.  Here we present the distribution of the global fine
structure infrared emission lines and constrain conditions in the
emitting environments over the full remnant.  The complex Doppler
velocity field as traced by these fine structure lines will be detailed
separately \citep{casa_doppler_inprep}.

\section{Observations and Reduction}
\label{sec:observ-reduct}

Low-resolution 5--38\um\ spectral maps were obtained in January, 2005
with the Infrared Spectograph \citep[IRS,][]{Houck2004a} aboard Spitzer.
A single mapping observation consisting of 364 individual rastered
spectra was obtained in the Long-Low module (LL, 15--38\um), utilizing
5.08\arcsec\ steps across the slit (1/2 slit width), and 158\arcsec\
steps along the slit (approximately the full sub-slit length), for a
total of 12.6 seconds per position.  Short-Low (SL, 5--15\um) coverage
was obtained by tiling four separate spectral maps to accomodate the
maximum observation duration limits of the IRS, comprising 1218 total
independent spectra.  Precise scheduling was required to fully tile the
remnant without allowing gaps to open between the quadrants.  The SL
slit was stepped at 2\arcsec$\times$52\arcsec, for slightly less than
double coverage and an effective exposure time of 11.4s per position.
The best effective resolution is $\sim$2\arcsec\ in SL and
$\sim$8\arcsec\ in LL (varying linearly with wavelength).  The total
angular coverages obtained were 11\farcm0$\times$7\farcm8 (LL) and
6\farcm3$\times$5\farcm9 (SL).

Background removal was performed \emph{in situ}, using spectral
pointings selected from the periphery of the maps, choosing regions with
low relative 24\um\ surface brightness.  Since the IRS sub-slits (mapped
to separate grating orders) are separated by 180\arcsec\ (LL) and
68\arcsec\ (SL), multiple positions outside the main remant are obtained
in the course of a full spectral map in both orders, which aids in
selecting suitable background spectra.  However, since Cas A lies close
to the Galactic plane, foreground and background emission from
intervening and background Galactic cirrus clouds --- including
molecular hydrogen, PAH emission, and weak low ionization emission lines
--- is present throughout the region.  This complex filamentary
emission, readily seen in extended 24\um\ maps \citep{Kim2008}, varies
considerably over the face of the remnant, and is also present in the
selected background regions.  Although the contaminating line emission
is very faint --- $\lesssim$2\% of the line surface brightness of the
bright emitting ring --- it provides a modest limitation on our ability
to recover fluxes for the lowest surface brightness lines (see also
Fig.~\ref{fig:sl_full}).

All IRS data were reduced from pipeline version S17 data products with
\textsc{Cubism}, a custom tool created for the assembly and analysis of
spatially-resolved spectral cubes from IRS spectral maps
\citep{2007PASP..119.1133S}.  A total of four separate spectral cubes
were created, one each for orders 1 and 2 of both SL and LL.  The flux
calibration was achieved using appropriate extended-source FLUXCON
corrections, and aberrant or ``rogue'' pixels were automatically cleaned
using the statistical methods built into \textsc{Cubism}, with a small
amount of additional manual removal.  The short 6.3s exposure ramps
contain only four samples, which leads to occasional anamolous cosmic
ray correction, inducing small bad pixel clusters.  These were mitigated
with statistical record-level bad pixel flagging.

\section{Spectral Maps}

\begin{figure*}
\plotone{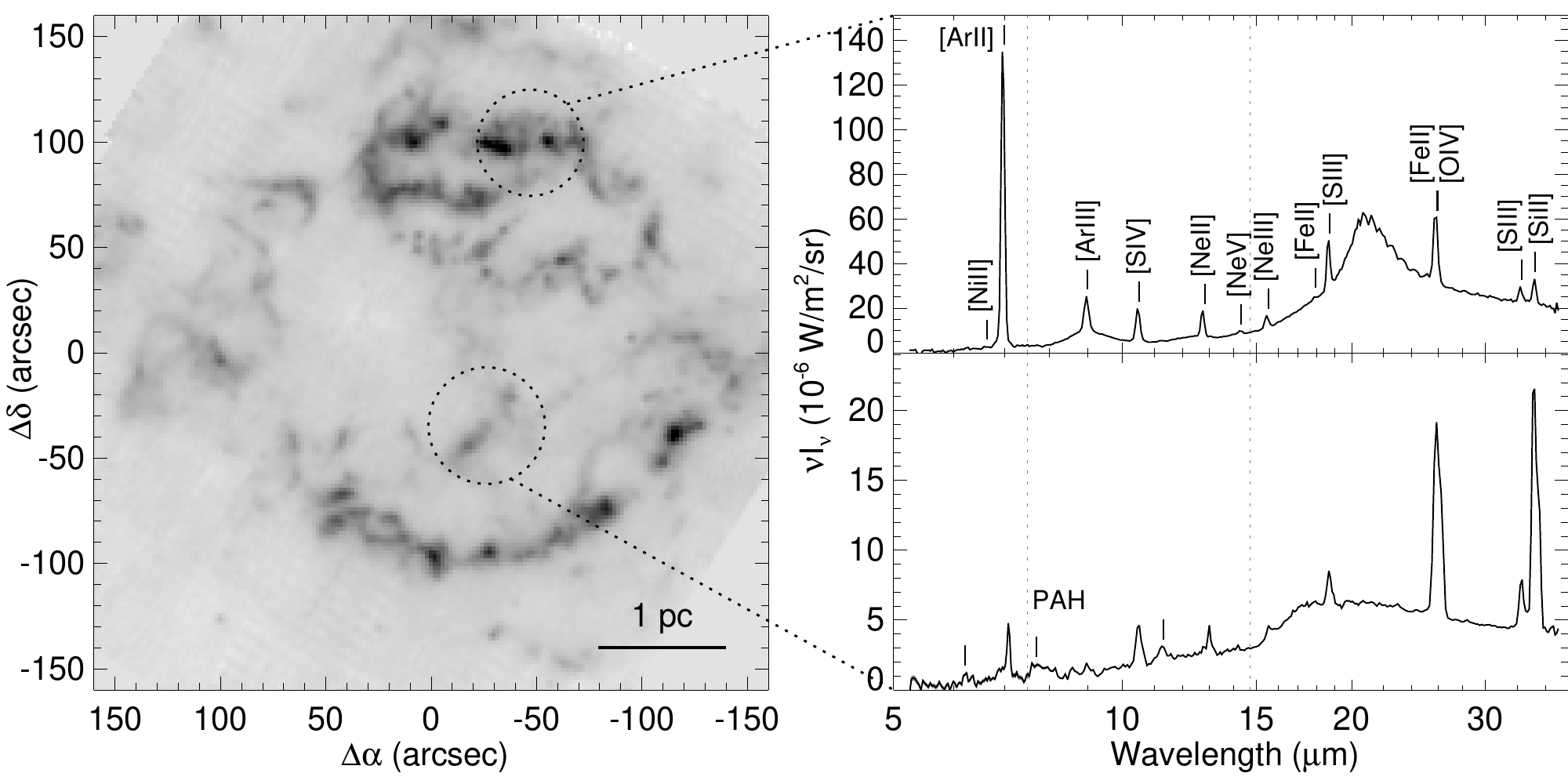}
\caption{At left, a full 7.5\um--14.7\um\ map created from the SL1 cube,
  320\arcsec$\times$320\arcsec, square root scaled, with offset
  positions labelled relative to the expansion center
  \citep[23:23:27.77, 58:48:49.4;][]{Thorstensen2001}.  Representative
  spectra extracted in circular apertures are shown at right.  Lines
  detected from the ejecta are labeled on the spectrum of the bright
  northern rim at top, and residual PAH features associated with the
  Galactic foreground are indicated on the much fainter interior
  emission spectrum below.  Broad peaks at 9 and 21\um\ are due to dust
  emission \citep[see][]{2008ApJ...673..271R}, and the periodic signal
  at 20--25\um\ is instrumental fringing.  The wavelength range used to
  construct the map is indicated with vertical dotted lines.}
\label{fig:sl_full}
\end{figure*}
 
\begin{deluxetable}{lcr@{\,$\pm$\,}lc}
  \tablecaption{Detected Emission Lines\label{tab:lines}}
  \tablecolumns{5}

  \tablehead{
    \colhead{Line} &
    \colhead{Wavelength} &
    \multicolumn{2}{c}{Peak Intensity} &
    \colhead{Total Flux}\\
    \colhead{} &
    \colhead{\um} &
    \multicolumn{2}{c}{$10^{-7}$\,W/m$^2$/sr} & 
    \colhead{$10^{-13}$\,W/m$^2$} \\
    \colhead{(1)} &
    \colhead{(2)} &
    \multicolumn{2}{c}{(3)} & 
    \colhead{(4)}}
\startdata
\niII  &  6.636 &    2.4 &  0.85 &  \nodata \\
\arII  &  6.985 &    221.&  0.86 &  7.72    \\
\arIII &  8.991 &   41.7 &  0.82 &  1.24    \\
\sIV   & 10.511 &   21.0 &  0.43 &  1.17    \\
\neII  & 12.814 &   33.7 &  0.21 &  1.41    \\
\neV   & 14.322 &   1.75 & 0.17  &  \nodata \\
\neIII & 15.555 &   8.62 &  0.17 &  0.49    \\
\feII  & 17.936 &   2.49  & 0.14 &  \nodata \\
\sIII  & 18.713 &   13.2 &  0.23 &  1.44    \\
\oIV\tablenotemark{a}   & 25.890 &   27.1 &  0.29 &  4.95      \\
\feII\tablenotemark{a}  & 25.988 &   \multicolumn{2}{c}{\nodata}  & \nodata \\
\sIII  & 33.481 &   4.08 &  0.18 &  1.02    \\
\siII  & 34.815 &   12.3 &  0.21 &  3.16   
\enddata
\tablecomments{Col. (1): line name.  Col. (2): rest wavelength.  Col.
  (3): peak surface brightness over 1.85\arcsec$^2$ (SL) and
  5.08\arcsec$^2$ (LL) areas.  Col (4): total flux integrated over the
  remnant in a 5\farcm3$\times$5\farcm3 region; statistical
  uncertainties in the total are 0.01 or less in these units. Faint
  lines excluded.} \tablenotetext{a}{\oIV\ and \feII\ are blended into a
  single emission peak at LL resolution.}
\end{deluxetable}


A full 7.5--14.7\um\ map created from SL order 1 is shown in
Fig.~\ref{fig:sl_full}, and offers excellent correspondence with the
8\um\ IRAC map of \citet{2006ApJ...652..376E}. Two representative full
spectra extracted from the interior and along the bright northern rim
are also shown.  These two spectra highlight the large dynamic range of
line intensity, line ratios, and continuum shape across the remnant
face.  The broad peaks at 9 and 21\um\ arise from warm protosilicate and
other dust emission \citep[see][]{2008ApJ...673..271R}.

\begin{figure*}
\plotone{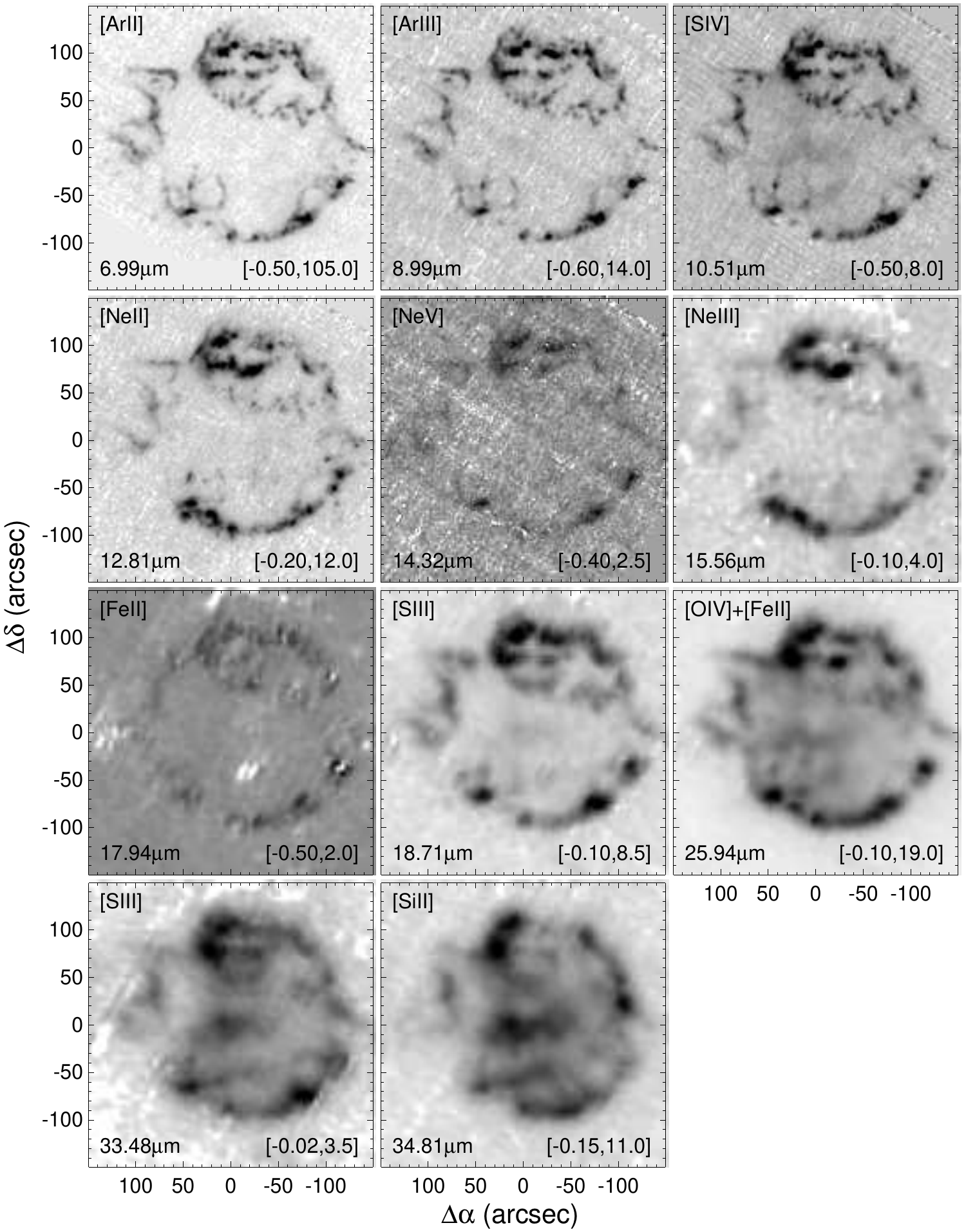}
\caption{Surface brightness maps in the eleven brightest emission lines,
  sorted by order of increasing wavelength.  Each image is
  300\arcsec$\times$300\arcsec, with offset positions as in
  Fig.~\ref{fig:sl_full}.  The integrated, continuum-subtracted line
  surface brightness is square root scaled between low and high
  thresholds specified in brackets in each pane (in units of
  $10^{-7}$\,W/m$^2$/sr).}
\label{fig:line_maps}
\end{figure*}

Surface brightness maps from the eleven brightest emission lines
(readily seen in Fig.~\ref{fig:sl_full}) are shown in
Fig.~\ref{fig:line_maps}.  Spectral line maps were created by
integrating the spectral cube intensities over a wavelength region large
enough to sample the line flux at all velocities present within the
remnant \citep[roughly -5000\,km/s to +7000\,km/s,
see][]{casa_doppler_inprep}.  Prior to integration, the local dust and
synchrotron continuum emission was removed by subtracting a weighted
average of flanking wavelength regions, carefully chosen to avoid
contamination by nearby lines at all velocities.  The averaging weight
used at each wavelength within the line region ($\lambda_{line}$) for a
given continuum wavelength ($\lambda_{cont}$) was
$|\lambda_{cont}-\lambda_{line}|^{-1}$.

The broad dust feature present throughout most of the ejecta has a sharp
peak at 21\um, which induces a steep and variable continuum in the
15--21\um\ range (see Fig.~\ref{fig:sl_full}).  Estimating accurate line
map fluxes in this range (LL order 2) required a preliminary continuum
removal step.  The cube was first smoothed spatially by three pixels.  A
linear or quadratic background function at each (smoothed) spatial
position was fitted using two to four wavelength intervals free of line
contamination, separately chosen for each line in the range.  This
noiseless background estimate was removed from the unsmoothed cube, and
flanking continuum regions were then subtracted as above to eliminate
any residual continuum contributions underlying the line.  For the faint
\feII\ 17.94\um\ line in this region, uncertainty in the continuum
subtraction dominates the residual systematic errors.

Table \ref{tab:lines} lists the lines detected at low resolution, with
the peak surface brightness along the bright ring, as well as integrated
line flux over the entire remnant.  Very faint lines known to be
associated primarily with Galactic cirrus emission, including rotational
H$_2$ S(0)--S(2) and PAH emission features, are omitted, as are the
integrated flux measurement of the faintest ejecta lines, which would be
dominated by systematic errors.  Among the brighter lines, systematic
uncertainties in the total flux are 10\% or less.  Signal to noise in
the brightest regions of the line maps was typically 30 or above.
\arII\ is by far the brightest line detected, and as it also offers the
best spatial resolution, has proven very useful for Doppler
decomposition of the remnant.

In many cases line identification is hampered by Doppler broadening and,
less frequently, multiple line-of-sight velocity components.  The \oIV\
line at 25.89\um\ is a special example of such a case, as it is
profoundly blended with \feII\ 25.99\um\ even at high spectral
resolution.  The line map at 26\um\ therefore contains contributions
from both of these lines.  Constraints on the fractional contribution of
\feII\ and \oIV\ in different regions can be made using expected line
ratios to other lines of Fe, inluding \feII\ 17.94\um, as well as
\feIII\ 22.93\um, detected in our follow-on high-resolution maps of
selected regions.  Such an analysis will be considered in a subsequent
study, but is consistent with both \feII\ and \oIV\ contributing and in
differing amounts across the remnant.

\begin{figure}
\plotone{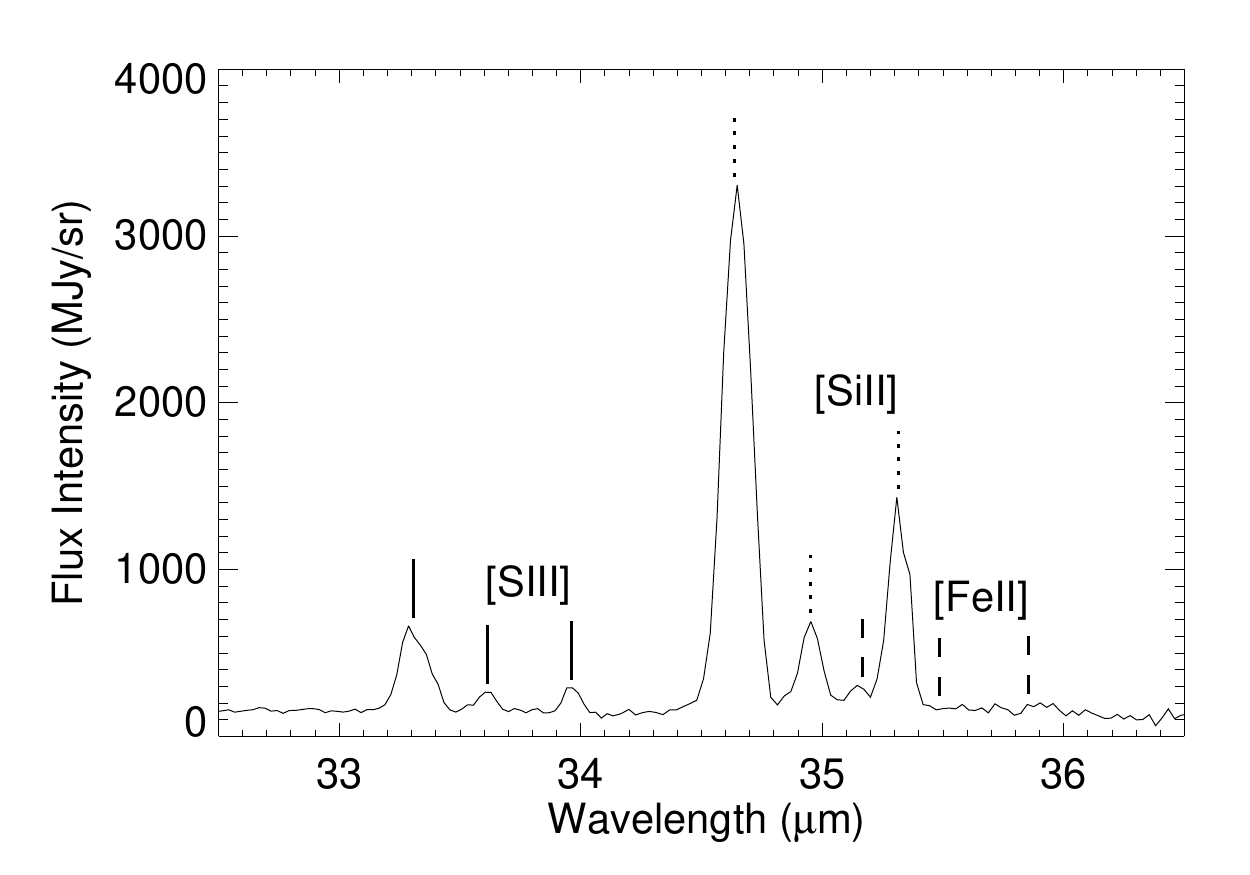}
\caption{A portion of the high-resolution IRS spectrum of the central
  remnant, used to identify \siII\ emission at 35.3\um.  The \sIII\
  33.48\um\ line was used to fit three velocity components at
  -1549\,km/s, 1175\,km/s, and 4303\,km/s.  These three components are
  marked for \sIII\ (solid lines), \siII\ (dotted), and a weak line of
  \feII\ not detected in our low resolution data set (dashed).}
\label{fig:highres}
\end{figure}

In cases where Doppler confusion could support more than one line
identification (most significant at the remnant center), we used our
high-resolution spectra to resolve the ambiguity.  An example for \siII\
is given in Fig.~\ref{fig:highres}.  A consistent set of three velocity
components is visible for \sIII, \siII, and a faint \feII\ line
undetected in our low-resolution data set.

A bright ring of emission, with overlapping smaller sub-rings most
clearly seen in the N and SW, is the dominant structure evident in all
line maps, but significant variations in the line strengths occur
throughout, in particular when comparing the interior to the ring.  For
the most part, the line strength variations on the bright ring are not
complicated by the three-dimensional structure of Cas A.  This is
apparent in the Doppler reconstruction of \citet{Lawrence1995} which
shows that there are very few overlapping structures along the same line
of sight, with the exception of the NNE and the base of the NE jet.  The
same cannot be said of the interior emission, where nearly every line of
sight has two or more dominant Doppler components
\citep{casa_doppler_inprep}.

\section{Progression of the Shock}
\label{sec:shock-progress}

\begin{figure}
\plotone{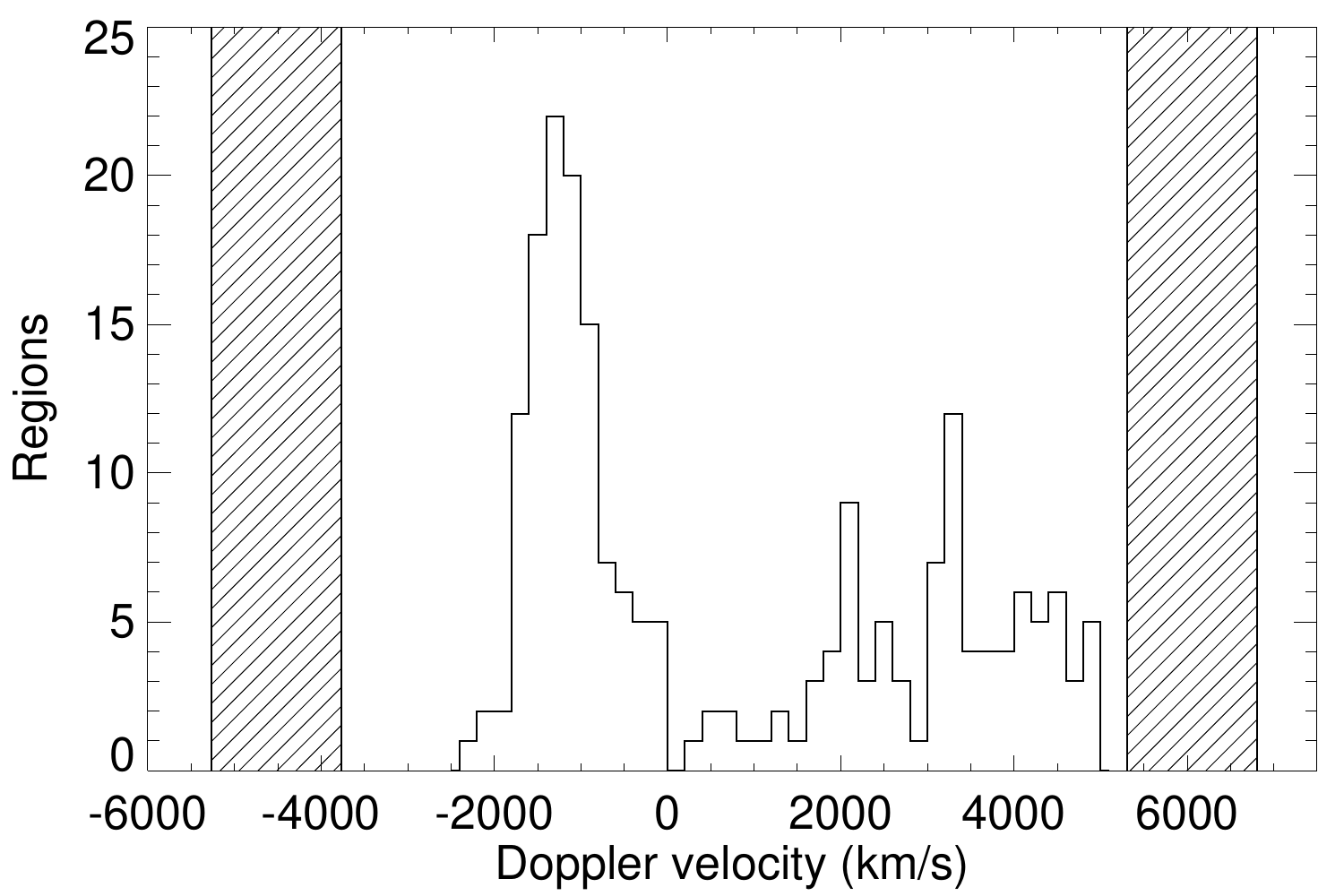}
\caption{Velocity distribution of \siII\ emission interior to the bright
  ring.  Hatched regions represent the expected velocities (-4520\,km/s,
  +6061\,km/s) and the velocity width (1500\,km/s) of the reverse shock
  at the remnant's center \citep{Reed1995}.}
\label{fig:siii_doppler}
\end{figure}

A reverse shock driven by the interaction of high velocity ejecta with
the circumstellar material is propagating backwards into the outflowing
material at a speed of 2500--4400\,km/s \citep[relative to the bulk
outflow velocity,][]{DeLaney2003,Morse2004}.  Material inside the
reverse shock boundary was ``once-shocked'' by the initial outgoing
blast wave, but has subsequently cooled and recombined, and has not yet
been re-heated and re-ionized by the returning reverse shock.  Three
dimensional Doppler reconstructions of optical and infrared lines such
as \arII\ show that they illuminate the reverse shock in a broad band
approximately in the plane of the sky, forming the bright ring
\citep{Lawrence1995,casa_doppler_inprep}.  There is little or no
emission from these ring lines across the face of the remnant; however,
interior to the ring, strong emission lines of the low ionization
species of S and Si appear.  The velocity distribution of the interior
\siII\ emission, shown in Fig.~\ref{fig:siii_doppler}, demonstrates that
this central material lies fully inside the reverse shock boundary.  In
addition to these lines, strong 26\um\ emission is present in the
interior, contributed by \feII\ and/or \oIV.

\subsection{Electron Density}
\label{sec:electron-density}

\begin{figure}
\plotone{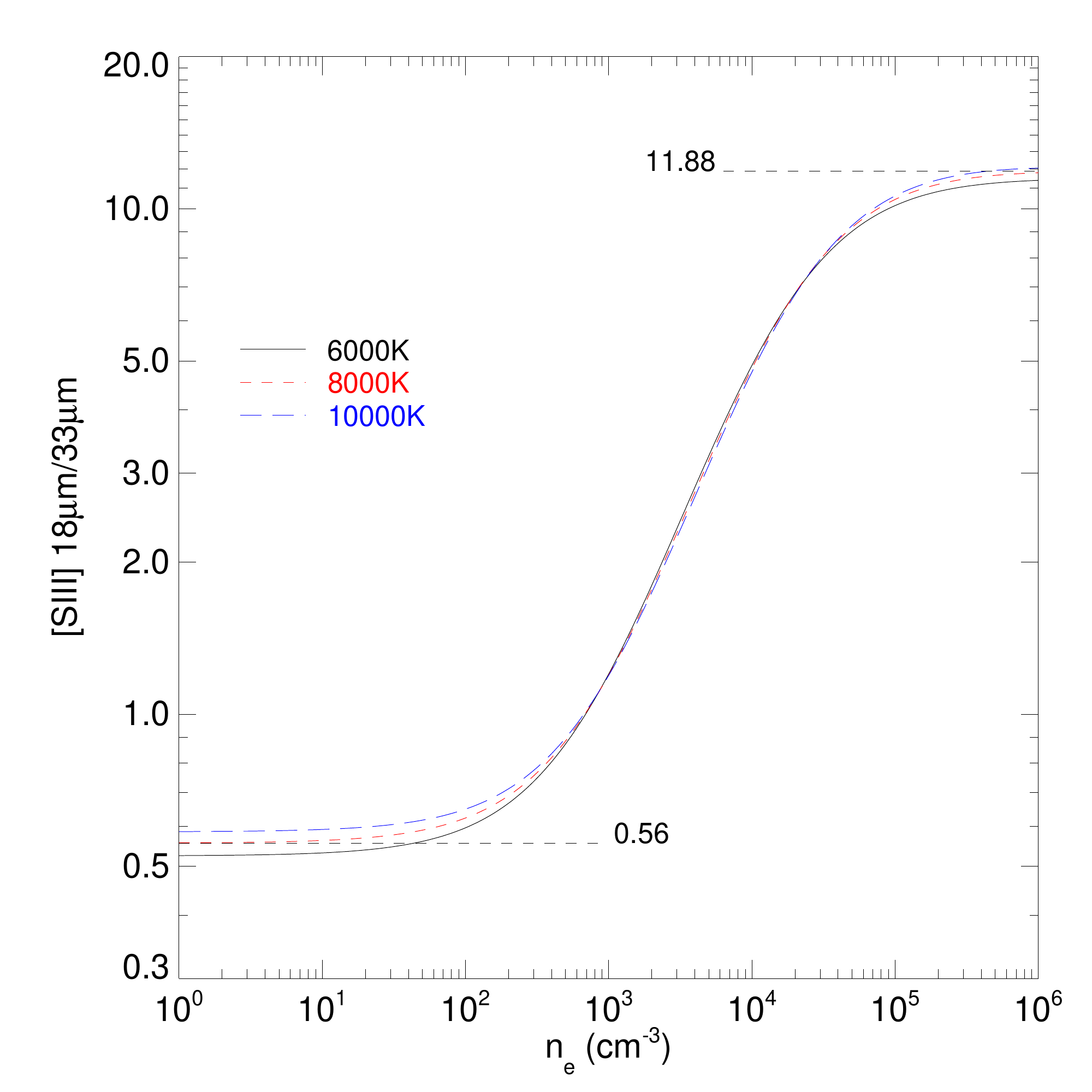}
\caption{The \sIII\,18\um/\sIII\,33\um\ density diagnostic for three
 representative temperatures.}
\label{fig:siii_density}
\end{figure}

\begin{figure}
\plotone{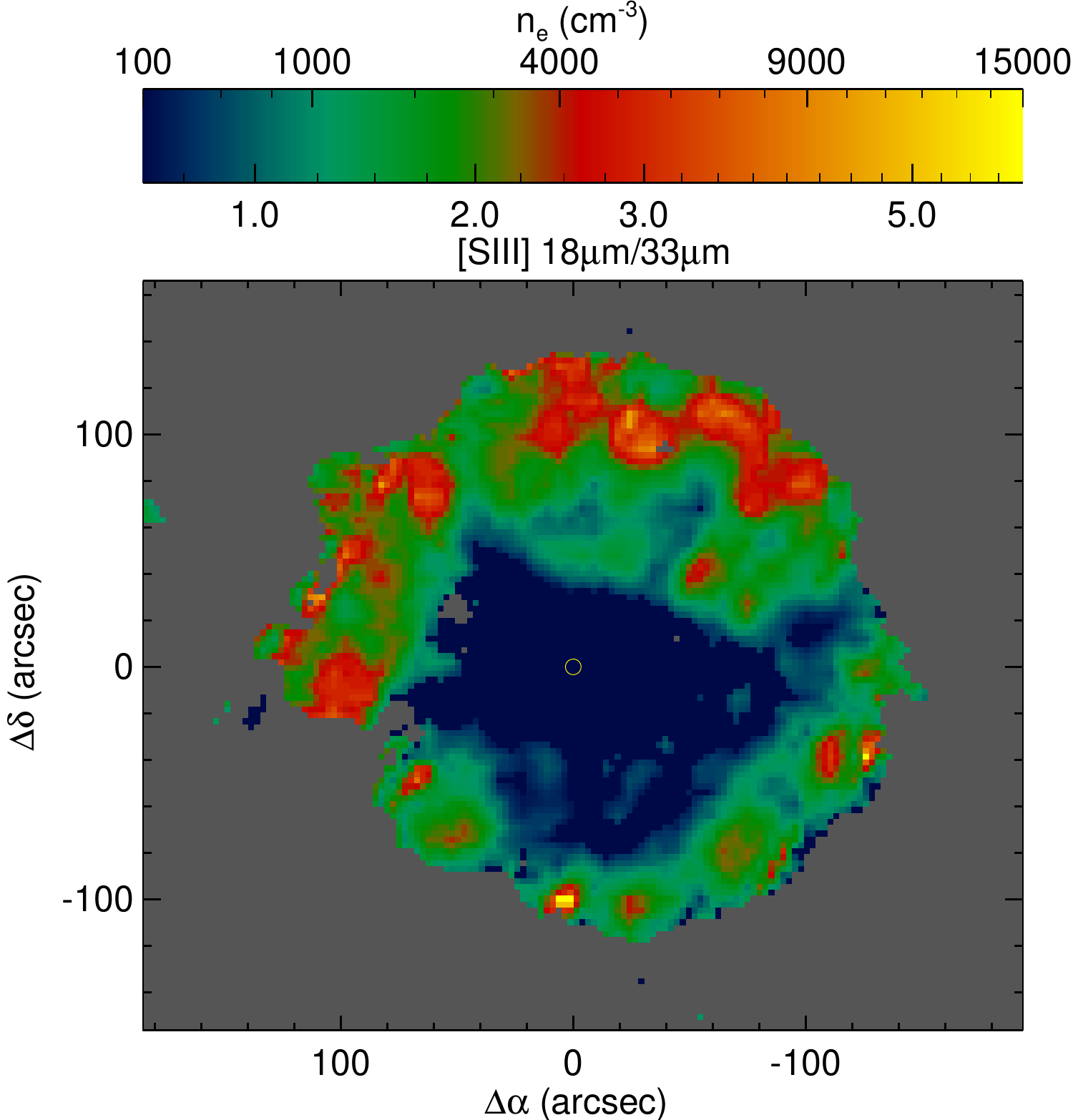}
\caption{The electron density in the remnant, computed from the
  \sIII\,18\um/\sIII\,33\um\ ratio.  Regions without sufficient signal
  in the ratio map are shown gray.  The darkest regions, which include
  most of the interior emission, are in the low density limit of
  $n_e\lesssim100$\,cm$^{-3}$.  The expansion center position of
  \citet{Thorstensen2001} is marked as a small circle, with offsets
  relative to it.}
\label{fig:siii_density_map}
\end{figure}

We can probe the change in density of the ejecta as they encounter the
reverse shock using the \sIII\ lines.  A temperature-insensitive density
diagnostic is available in the 18.71\um/33.48\um\ ground-state fine
structure doublet line ratio, arising from low-lying collisionally
excited levels.  Recovering electron density requires balancing the
rates of collisional excitation and de-excitation and radiative
transitions into and out of the relevant levels of the ion.  For ions
with p$^2$, p$^3$ and p$^4$ ground state electron configurations, like
\myion{S}{++}, it is sufficient to consider only the 5 lowest lying
energy levels \citep{Osterbrock2006}.  Up to date temperature-dependent
collisional rates were taken from the IRON project \citep{Hummer1993}
using its TIPBASE database interface.  Radiative rates were obtained
either from the IRON project, or, where available and reliable, from the
NIST Atomic Spectra
Database\footnote{\url{http://physics.nist.gov/PhysRefData/ASD/}}.  The
five coupled linear rate equations were solved numerically for all level
populations, imposing the constraint that the population fractions sum
to unity.  This solution is performed for a given temperature and
density, interpolating the temperature-dependent collisional rates over
$\log(T)$.  The resulting line ratio dependence on density is
illustrated in Fig.~\ref{fig:siii_density}.  The
\sIII\,18\um/\sIII\,33\um\ line ratio saturates in the high density
limit at 11.9, when collisions are rapid enough to populate the levels
in equilibrium according to their statistical weights, and at a low
density limit of 0.56, when radiative effects dominate.

To recover electron density in the remnant, the two \sIII\ line maps
were spatially registered and convolved to matching resolution using a
custom convolution kernel constructed from Spitzer STinyTim PSF models
in the manner of \citet{2008ApJ...682..336G}.  The matched images were
then ratioed, and corrected for A$_{V}$=5 magnitudes of extinction using
the interstellar extinction curve of \citet{Chiar2006} (which results in
a modest 10\% increase). All data in the ratio map with signal-to-noise
below 2.25 were discarded, and map regions within $1\sigma$ of the low
or high-density ratio limits were set to the limiting densities.

The \sIII\,18\um/\sIII\,33\um\ ratio ranges from 0.5 near the center to
a maximum of 7.5 on the bright rim.  The resulting projected density map
is shown in Fig.~\ref{fig:siii_density_map}.  This density is averaged
over spatial scales of 10\arcsec\ (0.16pc).  Although the bright ring is
not generally confused by overlapping Doppler components at this spatial
resolution \citep{Lawrence1995}, high-resolution optical and X-ray
images resolve structure out at 0.2--1$\arcsec$
\citep{Fesen2001,Hughes2000}, so even without projection effects, some
blending of structure is unavoidable.  In regions where two or more
strong emission components are present along the line of sight, which
includes most of the interior emission, the \sIII\ line ratio represents
the average emission-weighted electron density.  However, individual
line ratios calculated for red- and blue-shifted \sIII\ line sets in the
interior emission near the remnant center yielded consistent values of
$\sim$0.5.

The interior \sIII\ emission is uniformly in or near the low limiting
density of the \sIII\ diagnostic ($n_e\lesssim100$\,cm$^{-3}$), whereas
the electron density in gas compressed by the reverse shock is much
higher, peaking above 1.5$\times$10$^4$\,cm$^{-3}$.  This general result
holds even assuming much lower temperatures, down to 1000\,K.  Unshocked
interior emission at much lower densities has also been observed in the
type Ia supernova remnant SN1006 \citep[e.g.][]{Hamilton1988}, and were
originally identified in Cas A through free-free absorption of low
frequency radio emission by \citet{Kassim1995}.  \citet{Hurford1996}
obtain densities ranging from 5000--15000\,cm$^{-3}$ for a range of
optical knot compositions, consistent with \citet{Chevalier1979}, and in
good agreement with our results on the bright rim.

Since the density enhancement in strong non-radiative shocks is limited
to a factor of four, the strong density enhancement provides evidence
for rapid radiative cooling of the infrared emitting clumps after they
pass through the shock.  Alternatively, hidden dense clumps may be
entrained in the outflowing material, which are illuminated only as they
pass through the shock.  This could then represent an extension to
higher densities of the clumpiness apparent in the more diffuse
X-ray-emitting ejecta \citep{Morse2004}.

\subsection{Ionization State}
\label{sec:ionization}

The progression of the reverse shock into the ejecta can be observed not
only in the electron density, which increases by at least two orders of
magnitude, but also in the ionization state of the outflowing atomic
material as it encounters the reverse shock.

\begin{figure}
\plotone{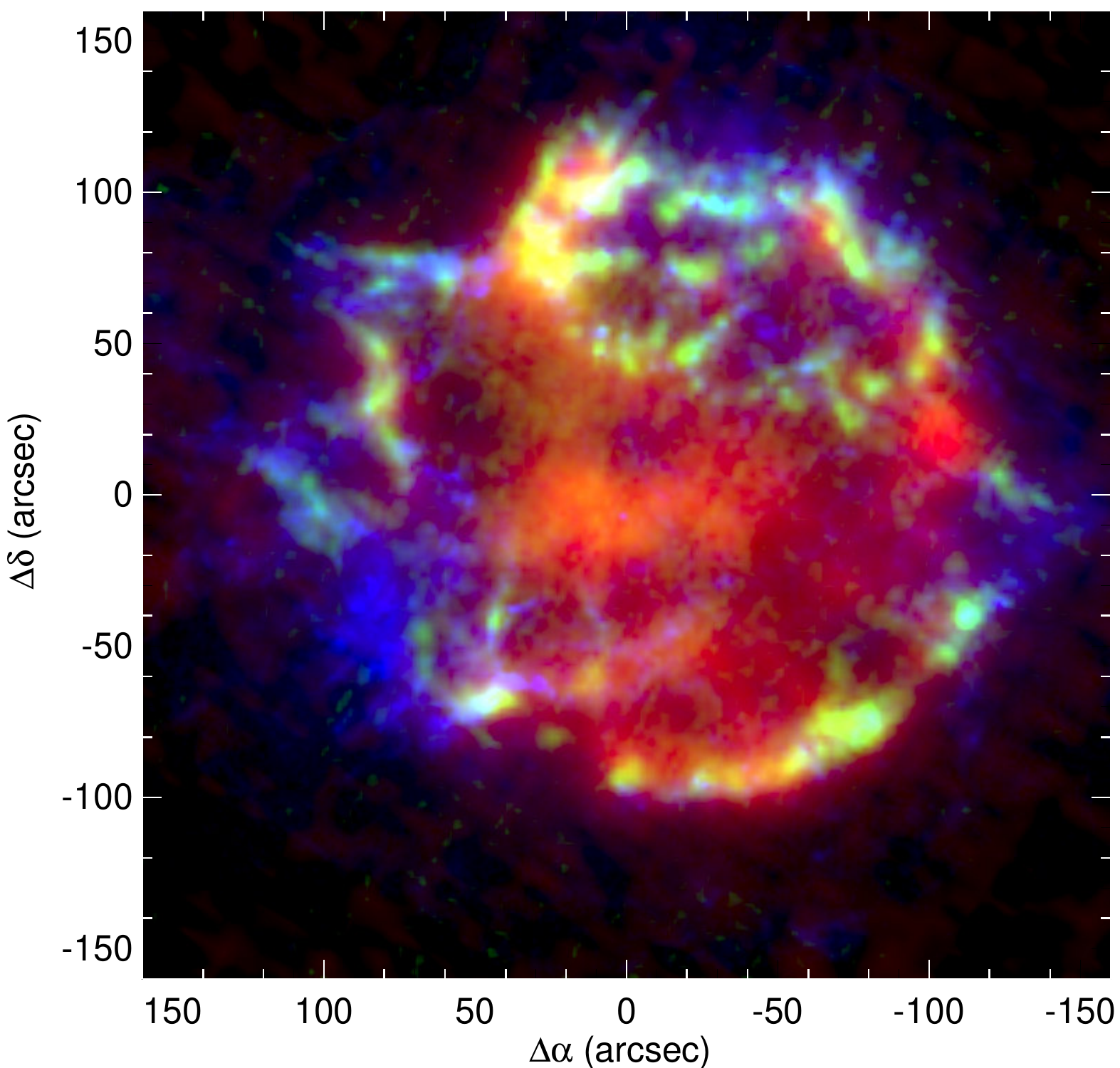}
\caption{A color coded map in \siII\ 34.8\um\ (red, ionization energy
  8.2eV), \sIV\ 10.5\um\ (green, ionization energy 34.8eV), and the
  summed X-ray emission of \siXIII+\siXIV\ (hydrogen and helium-like
  silicon, blue, ionization energy 0.5--2.4keV).  Square-root scaling is
  used for all three channels, and the coordinate scale follows
  Fig.~\ref{fig:line_maps}, with a 320\arcsec\ image size.  Immediately
  evident is the inside-out layering of ionization state, with low
  energy species in the unshocked interior region, and higher energy
  species layered within the non-uniform shock region itself.}

\label{fig:ionization}
\end{figure}

Figure~\ref{fig:ionization} illustrates this progression by comparing
line maps from two infrared lines arising from ions of differing
ionization potential --- \siII\ (8.2eV), and \sIV\ (34.8eV) --- together
with a summed pair of high-energy X-ray helium-like and hydrogen-like
K-alpha resonance lines: \siXIII\ (0.5keV), and \siXIV\ (2.4keV)
\citep{Hwang2000}.  There is clear layering of ionization state, from
low energy species in the interior, higher energies on the bright ring
where optical emission is also seen, and very high energies traced by
X-ray line emission extending beyond the infrared-bright rim itself.
Though line of sight projection of multiple structures complicates the
view in the northern rim, it is evident that the \emph{observed}
layering is highly non-uniform as a function of azimuth as well as along
the line of sight.

In various positions, more or less of the O-burning layer has
encountered the reverse shock.  In the direction along our line of
sight, very little of this material has reached the shock. These regions
are dominated by low ionization lines, \siII\ in particular (see
Fig.~\ref{fig:ionization}), which lie at lower velocities than the
shocked ring of emission (Fig.~\ref{fig:siii_doppler}).  In these
directions, neither element is seen in higher ionization states in the
infrared or X-ray.  This low-ionization pattern also appears at
different azimuths along the bright ring, indicative of positions where
O-burning ejecta moving approximately in the plane of the sky have yet
to encountered the reverse shock.  This emission is consistent with
material inside the reverse shock which has been ionized by ambient
Galactic stellar radiation and/or illumination from the rim of shocked
ejecta.

In other directions, moderate ionization state material traced by \sIV\
is present, with weak or absent X-ray emission.  Finally, there are
regions where only the high ionization Si X-ray lines are seen,
predominantly at large radii. These variations likely represent
different stages in the post-shock histories of the inhomogeneous
ejecta.  The low density X-ray filaments and diffuse material are
gradually heated and ionized as they penetrate the reverse shock, with a
time history that depends on the shape of the ejecta density profile
\citep{Laming2003}.  Higher density clumps of material can suffer a
different fate, reaching temperatures of only $\sim$10$^4$\,K as they
undergo rapid radiative cooling immediately post-shock, and are seen in
\sIV.

The detailed histories of shock-heated material requires considerably
more modeling, which incorporates variations in ejecta density and its
profile, ejecta velocities, and irregularities in the location of the
reverse shock.  It is not clear to what degree the density and
ionization state of the material we now see passing through the shock
front is determined by initial conditions of the ejecta --- diffuse
vs. clumpy, lower vs. higher velocity --- and to what degree the
heating, photo- and collisonal ionization, and subsequent cooling by
line radiation redistributes material between these phases.  What is
clear, however, is that the ejecta undergo a striking and rapid change
in density and ionization level as they encounter the reverse shock.

\section{Apparent Inhomogenous Mixing and the Neon Crescents}
\label{sec:neon-moons}

\begin{figure}
\plotone{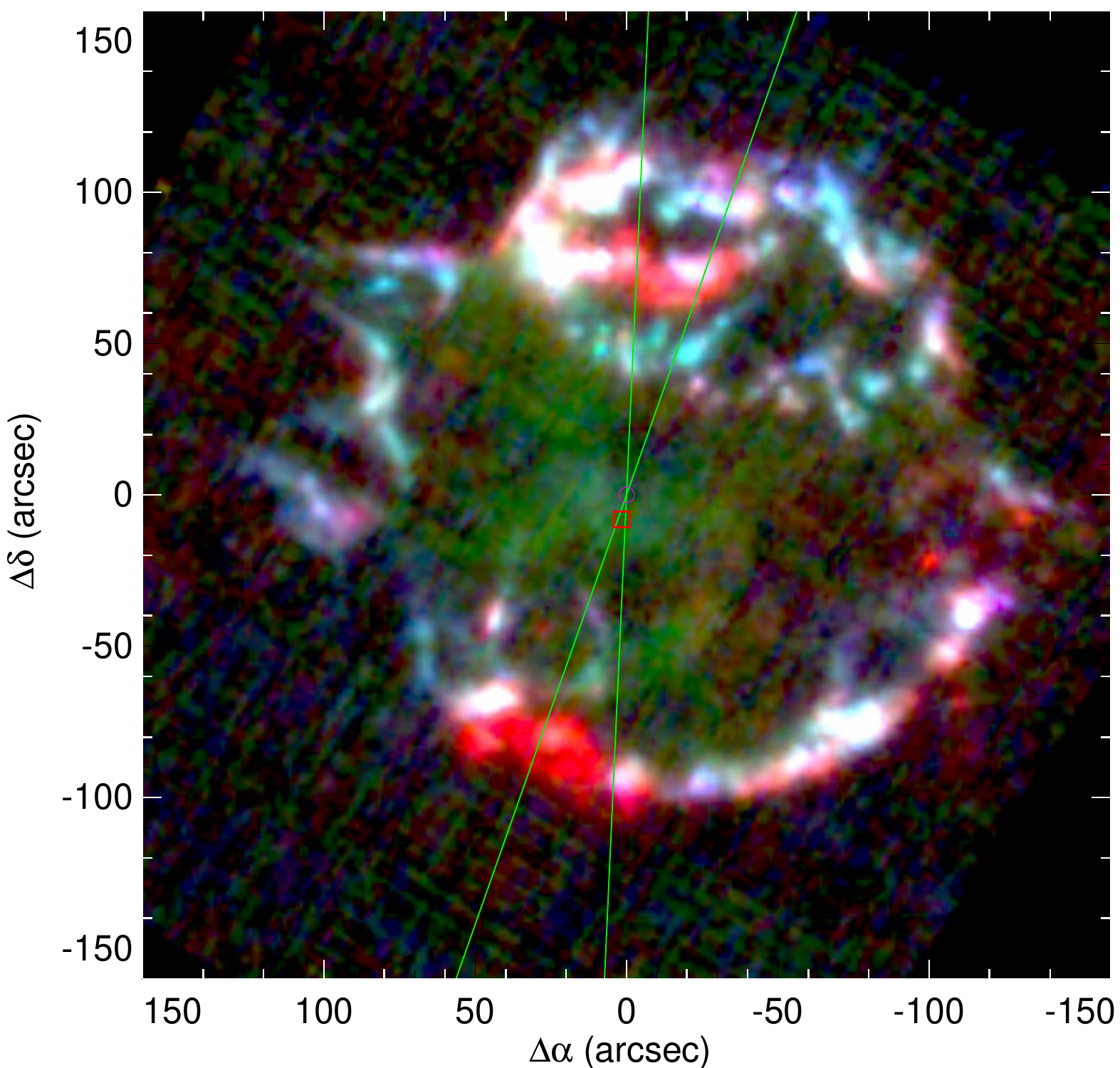}
\caption{Composite image of \arIII\ (blue), \sIV\ (green), and \neII\
  (red), showing two pronounced neon-rich crescent-shaped regions to the
  N and S (square root scaling).  Axes as
  Fig.~\ref{fig:ionization}. Also shown are the kinematic center of the
  remnant \citep[magenta circle,][]{Thorstensen2001} and X-ray
  localization of the remnant's compact object (red square, 7\arcsec\ to
  the south).  Green lines indicate the 1$\sigma$ range of the ``kick
  vector'' direction of the compact object from the ejecta's expansion
  center: 169$\degr\pm$8.4$\degr$ \citep{Fesen2006}.  The two regions of
  enhanced neon abundance lie very close to this projected direction.
  Several much smaller neon-enhanced regions lie in the West along the
  X-ray jet direction.}
\label{fig:neon_moons}
\end{figure}

The Cas A ejecta are by no means homogenous in composition or uniform in
their layering.  Using optical spectroscopy, \citet{Chevalier1979}
demonstrated abundance inhomogeneities in the oxygen burning products S,
Ar, and Ca among the outlying fast moving knots, and suggested differing
completeness levels of oxygen burning could explain these variations.
Macroscopic mixing of similar nucleosynthetic products was suggested by
\citet{Douvion1999} using ISO MIR line maps of Ar, Ne, and S covering a
portion of the bright rim.  \citet{Hughes2000} used Chandra X-ray
spectroscopy to identify spatial inversion of iron and oxygen burning
products.  It would be fair to say that \emph{non}-uniformity of
composition is the rule rather than the exception in Cas A.

In a stratified model in which different layers of ejecta material are
coming into contact with the reverse shock front in different
directions, the appearance of individual emission lines depends on 1)
the expansion velocity of material, 2) the position of the reverse shock
with respect to it, and 3) the subsequent ionization state of the
material as it passes through the shock.  In this context, even if the
gross nucleosynthetic layering is preserved in the outflow, with lighter
elements overlying heavier ones, differing outflow velocities in
different directions can generate an \emph{apparent} mixing of layers,
since different strata are now arriving at and passing through the
reverse shock boundary in different regions. Combined with line of sight
projection effects, this can profoundly complicate interpretation of
apparent elemental mixing, a conclusion also drawn in the jet-induced
explosion model of \citet{Wheeler2008}.

A powerful example of this apparent mixing is found in the neon maps.
Neon is among the most abundant elements in the carbon and oxygen
shell-burning cores of intermediate mass supernovae progenitors
\citep{Meakin2006}.  In contrast to the optical neon lines, which are
highly obscured by line of sight extinction, and therefore difficult to
study \citep[e.g.][]{Chevalier1979,Fesen1990}, the MIR neon lines are
bright and readily detected throughout the ejecta ring. 

A three color composite image of \arIII, \sIV, and \neII\ in Fig.
\ref{fig:neon_moons} reveals two bright, crescent-shaped clumps of
highly enhanced neon abundance to the North and South along the ring.
Lines of Ar, Si, and S are very faint in these regions, and they have
excess 4.5\um\ brightness \citep{2006ApJ...652..376E}.  The blended
complex \oIV+\feII\ is strong as well, though the coarser angular
resolution prevents completely separating the small regions from the
surrounding ring material at these longer wavelength.

The \neII/\arIII\ ratio changes dramatically --- from $0.8\pm0.4$ on the
rim outside of these crescent regions, to over 50 on the crescents
themselves.  Since the remnant regions occupied by the Ne crescents are
not overly confused by overlapping Doppler structures
\citep{Lawrence1995}, the large ratio differences represent real
abundance variations.  Variations in abundance from optical emission
lines of S and O are also observed at spatial resolutions of 1\arcsec\
\citep{Reed1995} and down to individual ejecta knots at the $0\farcs2$
level \citep{Fesen2001}.

Overlaid on the image are the remnant expansion center of
\citet{Thorstensen2001} as well as the location of the X-ray compact
object, which is offset from the kinematic center by $\sim$7\arcsec\ to
the South.  The projected direction of motion, representing the
asymmetric kick received by the compact object, is
169$\degr\pm$8.4$\degr$ \citep{Fesen2006}.  The symmetric neon crescents
lie nearly perfectly along this kick direction.  Though the true-space
velocity of relative motion between the compact object and kinematic
center is unknown, this strongly suggests a preferred axis of symmetry
in the outflow along this direction.

Of obvious interest is the total emitting mass of neon in the
crescents. The observed line flux $F$ of a collisionally excited
emission line is:

\begin{equation}
 \label{eq:1}
 F=\frac{h\nu A_{ul}}{4\pi d^2}\int_V f_{u} n_{\mathrm{ion}} dV
\end{equation}

\noindent where $A_{ul}$ is the radiative rate between the two levels
involved in the emission, $f_{u}$ is the fractional population of the
upper level, $n_{\mathrm{ion}}$ is the ion density, $V$ is the physical
volume over which the emission occurs, and $d$ is the distance.
Assuming the fractional density of the ion is constant over the emission
volume, the emitting mass of the ion is therefore

\begin{equation}
 \label{eq:2}
 M_\mathrm{ion}=m_\mathrm{ion}n_\mathrm{ion}V=\frac{4\pi d^2 F}{A_{ul}f_uh\nu}m_\mathrm{ion}
\end{equation}

\noindent where $m_\mathrm{ion}$ is the atomic mass.  We take a
representative electron density on the bright rim of
$n_e=$4500\,cm$^{-3}$ from the \sIII\ analysis of
\S\,\ref{sec:electron-density}, and bracket our uncertainty in the
temperature by adopting $T=$5500--8000\,K \citep[see][]{Hurford1996}.
Isolating the regions of elevated \neII/\arII\ using polygonal
apertures, we obtain the integrated fluxes for the crescents indicated
in Table~\ref{tab:ne-flux}.  Each crescent is approximately 0.25\,pc$^2$
in size, and contributes roughly 10\% of the remant's total \neII\ line
flux.  The \neIII\ line is also relatively strong in each crescent
region, but \neV\ is very faint and not seen in the Southern crescent.
The \neV/\neII\ ratio in the Northern crescent is $\gtrsim 8\times$ that
of the Southern region, although the velocity difference between the two
lines at that position is $\gtrsim 1000$\,km/s, suggesting that the
increased line of sight confusion in the Northern cap may play a role in
this discrepancy.  \myion{Ne}{4+} contributes negligibly to the mass.

\begin{deluxetable*}{lccr@{\,$\pm$\,}lr@{\,$\pm$\,}lr@{\,$\pm$\,}lr}
 \tablecaption{Neon Crescent Line Fluxes\label{tab:ne-flux}}
 \tablecolumns{7}
 \tablehead{
   \colhead{Name} &
   \colhead{Position} &
   \colhead{Area} &
   \multicolumn{2}{c}{$F_{\mathrm{\neII}}$\,\tablenotemark{b}} &
   \multicolumn{2}{c}{$F_{\mathrm{\neIII}}$\,\tablenotemark{b}} &
   \multicolumn{2}{c}{$F_{\mathrm{\neV}}$\,\tablenotemark{b}} &
   \colhead{M$_{\mathrm{Ne}}$\tablenotemark{a}} \\
   \colhead{} &
   \colhead{J2000} &
   \colhead{pc$^2$} &
   \multicolumn{6}{c}{$10^{-15}$\,W/m$^2$} &
   \colhead{$10^{-5}$\,M$_{\sun}$}}
\startdata
Ne South & 23:23:31.53 +58:47:22.6 & 0.33 & 15.9&0.04  &  5.59&0.06 & \multicolumn{2}{c}{$<$0.09\tablenotemark{c}} & 8.9\,$\pm$\,0.4\\ 
Ne North & 23:23:25.88 +58:50:07.2 & 0.21 & 16.9&0.04  &  5.62&0.05 & 0.735&0.02  & 9.7\,$\pm$\,0.5
\enddata
\tablenotetext{a}{Excluding neutral neon.}  \tablenotetext{b}{Corrected
  for $A_V=5$ magnitudes of extinction.  Statistical uncertainties only.
  Additional systematic uncertainties apply ($\sim$10\% for
  \neII,\neIII; $\sim$25\% for \neV).}  \tablenotetext{c}{3-$\sigma$
  upper limit}
\end{deluxetable*}

Assuming they emit in similar volumes, the densities of \myion{Ne}{++}
and \myion{Ne}{4+} relative to \myion{Ne}{+} can readily be estimated by
forming a ratio of Eq.~\ref{eq:1} for each of the higher ionized species
to the lower, which eliminates all quantities except the line flux,
upper level population, radiative rate, and frequency.  We solve for the
level populations using atomic data for neon obtained and applied in the
manner described in \S\,\ref{sec:electron-density}. This results in
$n_{\myion{Ne}{++}}/n_{\myion{Ne}{+}}=0.17$ in both regions, and, in the
North, $n_{\myion{Ne}{4+}}/n_{\myion{Ne}{+}}=0.003$.  Assuming these
ions are the dominant source of electrons, and taking the fractional
density of the unseen ion \myion{Ne}{+++} (which does not emit in the
MIR) as 0.05, we can estimate

\begin{equation}
 \label{eq:3}
 \frac{n_e}{n_{\myion{Ne}{+}}}=1+2\frac{n_{\myion{Ne}{++}}}{n_{\myion{Ne}{+}}}+3\frac{n_{\myion{Ne}{+++}}}{n_{\myion{Ne}{+}}}+4\frac{n_{\myion{Ne}{4+}}}{n_{\myion{Ne}{+}}}\approx 1.5
\end{equation}

Using Eq.~\ref{eq:2} to calculate the mass, we find roughly
1.8$\times10^{-4}$\,M$_\sun$ of neon in the crescent regions (dominated
by \myion{Ne}{+} and \myion{Ne}{++}, and excluding neutral neon).  This
mass is comparable to that found by \citet{Arendt1999} in other (larger)
segments of the rim.  The entire ionized neon mass of the remnant,
assuming an approximately constant density throughout the bright rim, is
8.6$\times10^{-4}$\,M$_\sun$, such that roughly 20\% of the neon mass is
concentrated in two small regions covering $\lesssim$2\% of the rim's
area.

The relative abundance of Ne to Ar (neglecting neutral species of both)
can be calculated using line maps of \arII\ and \arIII.  We assume the
same temperature and densities used for neon, and calculate level
populations as above.  The relative abundance increases from
$n_{\mathrm{Ne}}/n_{\mathrm{Ar}}$=2.7 averaged over the full remnant to
$n_{\mathrm{Ne}}/n_{\mathrm{Ar}}$=26 on the S crescent region.  These
likely represent lower limits on the full relative abundance, including
neutral species, as the first ionization potential of Ne (21.6\,eV) is
substantially larger than that of Ar (15.8\,eV), so that the average
neutral fraction can be assumed to be larger in the former.

Flowing outwards $\sim$20 degrees from the plane of the sky at roughly
-5500\,km/s (South) and +4200\,km/s \citep[North,
see][]{casa_doppler_inprep}, the ionized neon moving in these
crescent-shaped clumps represent only $4\times10^{-5}$ of the estimated
$10^{44}$\,J of expansion energy in the entire remnant
\citep{Willingale2003}.  Assuming the neutron star is moving in the
plane of the sky, the 7\farcs0 separation it has obtained from the
expansion center since the explosion \citep{Fesen2006} implies a
velocity of 342\,km/s.  Further assuming it to have a mass of
$\sim$1.4M$_\sun$ \citep[e.g.][]{Young2006}, the momentum impulse
imparted by the kick to the neutron star is then $\gtrsim 1000\times$
larger than that of either neon crescent.

Since they represent such small fractions of the full ejecta's kinetic
energy and the momentum of the neutron star, it is unlikely that even a
very large asymmetry in the neon crescents' initial velocities could
have directly contributed to the kick experienced by the compact object
with respect to the expanding ejecta.  In many ways, however, the
infrared emitting mass of neon in the crescents must be a significant
underestimate of the full column of material moving along this symmetric
pair of directions, with layers of heavier elements which have not yet
reached the reverse shock still largely unilluminated, neutral species
missing from the estimate, hotter X-ray gas at greater radii, and other
elements such as oxygen entrained in the crescents themselves not
readily accounted for.

In the final shell burning nucleosynthetic structure, the neon layer
lies above the heavier silicon, sulfur, and argon layers
\citep{Woosley1995}. If the gross layering of pre-explosion elements is
preserved, the strong overabundance of neon in these crescents along the
direction of the kick vector could in fact represent material which is
\emph{late} in meeting the reverse shock, having been accelerated to
slightly slower initial speeds.  If so, it could be expected that the
neon will gradually become less dominant as it progresses further beyond
the shock and cools, with sulfur and argon filling in behind it.  In
this sense, this pair of directions is intermediate between the bright
rim, cooling dominantly in lines of Ar, and the front and back surfaces
along the line of sight, along which very little material has yet
encountered the reverse shock.

The fact that the compact neon crescents are nearly equal in size and in
mass, and are arranged symmetrically about the expansion center,
strongly indicates they arise not from macroscopic or turbulent mixing
of the ejecta layers during expansion, but from a preferred axis in the
initial outflow configuration along this direction.

\section{Conclusions}
\label{sec:conclusions}

We present global low-resolution 5--37\um\ spectral maps of the
Cassiopeia A supernova remnant obtained with Spitzer's IRS spectrograph.
We detect numerous fine-structure emission lines arising from ions with
ionization energies of 8--100\,eV, and strengths that vary considerably
across the remnant.  The observed ions represent some of the most
abundant species in oxygen and carbon shell burning supernova progenitor
cores.  We find:

\begin{enumerate}

\item Strong interior emission of predominantly low ionization, heavy
  ionic species, dominated by \siII.  These regions lie inside the
  bright rim along the line of sight, and probe slower moving material
  which has cooled from the passage of the initial outgoing forward
  shock wave, but which has not yet encountered the reverse shock.

\item The electron density in the remnant increases rapidly from
  $\lesssim$\,100\,cm$^{-3}$ in the unshocked interior to
  $\sim\!10^4$\,cm$^{-3}$ on the bright ring of emission, indicative of
  rapid cooling of material passing through the shock front, or
  significant pre-shock density inhomogeneity.

\item The ionization state increases rapidly as the ejecta encounter the
  reverse shock, with low ionization species dominating inside the
  shock, moderate ionization species strongest just outside the shock,
  and very high ionization material seen in X-rays at the greatest
  distances.  These variations in post-shock history may depend on the
  density structure of the inhomogenous ejecta.

\item Two symmetrically arranged, compact, crescent-shaped clumps of
  material with highly enhanced neon abundance ($>10\times$ the remnant
  average) lie along the projected direction of the ``kick vector''
  separating the compact object from the expansion center of the
  remnant, tracing a new axis of explosion asymmetry in the ejecta.
  Since neon is expected to arrive at the reverse shock before heavier
  elements like Si and S, these could represent directions of
  \emph{reduced} initial velocity, intermediate between the bright rim,
  and the (still) quiescent material flowing towards the front and back
  faces of the remnant, which has yet to reach the reverse shock.

\item Since the ionization state and density vary steeply with distance
  from the expansion center, and have such a strong impact on the
  observed strength of individual emission lines, the apparent
  macroscopic mixing of elements in Cas A may result from variations in
  initial outflow velocity rather than bulk turbulent redistribution of
  the ejecta strata.  The symmetry in mass and outflow direction of the
  neon crescent regions strongly supports this interpretation.

\end{enumerate}

\acknowledgments 

The authors thank D. Arnett for helpful discussion, and an anonymous
referee for useful comments which improved this manuscript.  This work
is based on observations made with the Spitzer Space Telescope, which is
operated by the Jet Propulsion Laboratory, California Institute of
Technology under NASA contract 1407. Support for this work was provided
by NASA/JPL through award \#1264030.

\bibliographystyle{apj}
\bibliography{general,myref,inprep}

\end{document}